\documentclass[aps,prl,twocolumn,groupedaddress,showpacs,floatfix,superscriptaddress]{revtex4-1}
\usepackage{epsfig}
\usepackage{amsmath,amssymb}
\usepackage{graphicx}
\usepackage[dvipsnames,usenames]{color}
\usepackage[normalem]{ulem}
\usepackage{soul} 
\emergencystretch=\maxdimen
\begin{document}

\title{Langevin Simulations of the Half-Filled Cubic Holstein Model}
\author{B. Cohen-Stead} 
\affiliation{Department of Physics, University of California, 
Davis, CA 95616, USA}
\author{Kipton Barros}
\affiliation{Los Alamos National Laboratory, Physics and Chemistry of
Materials, Theoretical Division, Los Alamos, 87545}
\author{Zi Yang Meng}
\affiliation{Beijing National Laboratory for Condensed Matter Physics
and Institute of Physics, Chinese Academy of Sciences, Beijing 100190,
China}
\affiliation{Department of Physics and HKU-UCAS Joint Institute of Theoretical and Computational Physics, The University of Hong Kong, Pokfulam Road, Hong Kong SAR, China}
\affiliation{Songshan Lake Materials Laboratory, Dongguan, Guangdong 523808, China}
\author{Chuang Chen}
\affiliation{Beijing National Laboratory for Condensed Matter Physics
and Institute of Physics, Chinese Academy of Sciences, Beijing 100190,
China}
\author{R.T. Scalettar}
\affiliation{Department of Physics, University of California, 
Davis, CA 95616, USA}
\author{G. G. Batrouni}
\affiliation{Universit\'e C\^ote d'Azur, INPHYNI, CNRS, 0600 Nice,
France}
\affiliation{MajuLab, CNRS-UCA-SU-NUS-NTU International Joint Research
  Unit, 117542 Singapore}
\affiliation{Centre for Quantum Technologies, National University of
  Singapore, 2 Science Drive 3, 117542 Singapore}
\affiliation{Department of Physics, National University of Singapore, 2
  Science Drive 3, 117542 Singapore}
\affiliation{Beijing Computational Science Research Center, Beijing,
100193, China}

\begin{abstract}
Over the past several years, reliable Quantum Monte Carlo results for
the charge density wave transition temperature $T_{\rm cdw}$ of the
half-filled two dimensional Holstein model in square and honeycomb
lattices have become available for the first time.  Exploiting the further
development of numerical methodology, here we present results in three
dimensions, which are made possible through the use
of Langevin evolution of the quantum phonon degrees of freedom.  
In addition to determining $T_{\rm cdw}$ from
the scaling of the charge correlations, we also examine 
the nature of
charge order at general wave vectors for different temperatures,
couplings, and phonon frequencies, and the behavior of the spectral function
and specific heat.
\end{abstract}

\date{\today}

\maketitle


\noindent
\underbar{Introduction.}
Substantial effort has been devoted to developing and using Quantum Monte Carlo (QMC)
techniques to study the physics of interacting electrons.  
Auxiliary field methods formulated in real space, like Determinant
Quantum Monte Carlo (DQMC)~\cite{blankenbecler81,white89a,sorella89}, can determine correlations on clusters
of several hundreds of sites. 
However, unbiased approaches to studying electron correlations, such as DQMC, can be severely limited by the sign problem~\cite{loh90,iglovikov15},
unless additional constraints are imposed~\cite{zhang97a}.
The Dynamic Cluster Approximation~\cite{maier05} and Cluster Dynamical
Mean Field Theory~\cite{capone07,gull07} generalize single site Dynamical Mean
Field theory~\cite{metzner89,jarrell92,georges92,georges96,jarrell01,kyung06a} 
to finer momentum grids and generally have
a more benign sign problem than DQMC, 
allowing them to access lower temperatures
and/or more complex (e.g.~multi-band) models.
Diagrammatic QMC is another relatively new technology which 
is currently being developed~\cite{kozik10,kozik13}.
Despite the numerical challenges,
QMC applied to models with electron-electron interactions, like
the Hubbard model, has resulted in considerable qualitative 
insight into phenomena such as
the Mott transition, magnetic order, and, to a somewhat lesser extent,
exotic superconductivity (SC)~\cite{scalapino94} which arise from 
{\it electron-electron} interactions in real materials~\cite{dagotto05}.

Analogous strong correlation effects can
arise in solids due to {\it electron-phonon} coupling,  
including SC and charge density wave (CDW) formation; this
is the type of interaction we examine in this paper.
A simple model where such effects
can be studied is the Holstein Hamiltonian~\cite{holstein59}.
Early QMC work in two dimensions near half-filling
~\cite{scalettar89,noack91,vekic92,niyaz93,marsiglio90,hohenadler04}
examined CDW formation and its competition with SC.
A second generation of simulations has considerably improved
the quantitative accuracy of results looking at both finite temperature~\cite{weber18,chuang18,cohenstead19}
and quantum critical point~\cite{chen19,zhang19} physics
in two spatial dimensions on square and honeycomb lattices. 
Much of this progress has been possible thanks to newer QMC methods such as continuous 
time~\cite{weber18} and self-learning Monte Carlo~\cite{XYXuSelf2017,chuang18}. 
However, despite these improvements in effective update
schemes, the cubic scaling with lattice size $N$ of 
real space QMC methods
employed in existing work
has precluded similar studies in three dimensions.

We report here QMC simulations of the half-filled Holstein model
on cubic lattices as large as $N=14^3$ sites.  
These studies are made possible by employing a linear-scaling 
QMC method based on a Langevin evolution of the phonon degrees of
freedom~\cite{batrouni19a,batrouni19b,HubbardHolstein,HMC}. 
The large linear sizes that are accessible allow us to
perform the finite size scaling needed to extract the CDW
transition temperature $T_{\rm cdw}$ and also obtain the
momentum dependence of the charge structure factor $S({\bf k})$
to reasonable resolution.
We supplement the extraction of $T_{\rm cdw}$ from
$S_{\rm cdw} \equiv S(\pi,\pi,\pi)$ 
with calculation of the specific heat and spectral
function, and show that, while they provide a less precise determination
of $T_{\rm cdw}$, their features are consistent with those obtained from
$S_{\rm cdw}$.


\noindent
\underbar{Model and Methods.}
The Holstein Hamiltonian, 
\begin{align} \label{eq:Holst_hamil}
\nonumber \mathcal{\hat H} = & -t \sum_{\langle \mathbf{i}, \mathbf{j}
  \rangle, \sigma} \big(\hat c^{\dagger}_{\mathbf{i} \sigma} \hat
c^{\phantom{\dagger}}_{\mathbf{j} \sigma} + {\rm h.c.} \big) - \mu
\sum_{\mathbf{i}, \sigma} \hat n_{\mathbf{i}, \sigma} \\ & +
\frac{1}{2} \sum_{ \mathbf{i} } \hat{P}^{2}_{\mathbf{i}} +
\frac{\omega_{\, 0}^{2}}{2} \sum_{ \mathbf{i} }
\hat{X}^{2}_{\mathbf{i}} + \lambda \sum_{\mathbf{i}, \sigma} \hat
n_{\mathbf{i}, \sigma} \hat{X}_{\mathbf{i}} \,\,,
\end{align}
describes the coupling of electrons, with
creation and destruction operators
$\hat c^{\dagger}_{\mathbf{i} \sigma},
\hat c^{\phantom{\dagger}}_{\mathbf{i} \sigma}$,
to dispersionless phonon degrees of
freedom $\hat P_{\mathbf{i}}, \hat X_{\mathbf{i}}$, with the phonon mass normalized to $M=1$.
The parameter $t$ multiplies a near-neighbor
hopping (kinetic energy) term.  We set $t=1$ as our unit of energy,
resulting in an electronic bandwidth for the cubic
lattice equal to $W=12$. The coupling between the phonon
displacement and electron density on site $\mathbf{i}$ is controlled by $\lambda$ 
while the chemical potential, $\mu$, tunes the filling. 
In this study we focus
on half-filling, obtained by setting
$\mu=-\lambda^2/\omega_0^2$,
and report results in terms of a dimensionless electron-phonon coupling constant
$\lambda_D = \lambda^2/(\omega_0^2 W)$.
Despite its simplifications, the Holstein model
captures many qualitative features of
electron-phonon physics, including polaronic effects in the dilute limit~\cite{romero99,ku02,marchand13},
SC and CDW formation, and their competition~\cite{scalettar89b,aubry89,zheng97,grzybowski07,esterlis18,weber18,chen19,zhang19}.

The fermionic degrees of freedom appear only quadratically in the
Holstein model, Eq.~\eqref{eq:Holst_hamil}. Consequently, the fermions can
be ``integrated out" resulting in the product of two identical matrix determinants which are
nontrivial functions of the space and imaginary time dependent phonon
field.  The product of the two identical determinants is positive; thus 
there is no sign problem.  Most prior numerical
studies of the Holstein model employed DQMC,
which explicitly calculates changes in the determinant as the phonon
field is updated.  At fixed temperature, DQMC scales cubically in the number of sites $N$, and hence as $L^9$, where $L$ is the linear system size in 3D.  This
limits DQMC simulations in three dimensions to relatively small $L$.

Instead, we use a method based on Langevin updates which
exhibits linear near scaling in $N$.  Such methods were first formulated
for lattice gauge theories~\cite{batrouni85,davies86,batrouni87}. Attempts to simulate the Hubbard Hamiltonian with Langevin updates were limited to relatively weak coupling and high temperature by the ill-conditioned nature of the matrices, due to rapid fluctuations of the sampled Hubbard-Stratonvich fields in the imaginary time direction~\cite{scalettar86}. However, in the Holstein model the sampled phonon fields have an associated kinetic energy cost that moderates these fluctuations, giving rise to better conditioned matrices.

Here we briefly discuss the key steps in the algorithm and leave the details to Refs.~\cite{batrouni19a,batrouni19b}.
The partition function for the Holstein model is
first expressed as a path integral in the phonon coordinates,
$x({\bf i},\tau)$, by discretizing the inverse temperature
$\beta = L_{\tau} \Delta \tau$.  After performing the trace 
over the fermion coordinates, the phonon action ${\cal S}$ includes a term
${\rm ln}\,({\rm det} {\cal M})$ where ${\cal M}$ is a matrix of dimension
$NL_\tau$.  The phonon field is then evolved in a fictitious Langevin time
$t$ with $x({\bf i},\tau,t)$ moving under a force 
$\partial {\cal S} / \partial{x({\bf i},\tau,t)}$ 
and a stochastic noise term.  The part of the derivative of ${\cal S}$ 
which involves ${\rm ln}\,({\rm det} {\cal M})$ is evaluated with a
stochastic estimator.  It is necessary to compute ${\cal M}^{-1}$ 
acting on vectors of length $NL_{\tau}$, which is done using the conjugate
gradient (CG) method. An essential refinement of the algorithm is the application of Fourier Acceleration~\cite{batrouni85,davies86,batrouni87} to reduce critical slowing down resulting from the slow phonon dynamics in imaginary time.


Elements of the fermionic Green function are also obtained with
a stochastic estimator.  Once evaluated, one can measure all physical
observables.
We focus here on the charge structure factor,
\begin{align}
S({\bf k}) &\equiv  \sum_{\bf r}
c({\bf r}) \,  
e^{i {\bf k} \cdot {\bf r}},
\nonumber \\
c({\bf r}) &= 
\langle \, n_{\bf j+r} n_{\bf j} \, \rangle,
\label{eq:Sandf}
\end{align}
($n_{\bf j} = n_{{\bf j}\uparrow} + n_{{\bf j}\downarrow}$),
and the specific heat $C=d\langle E \rangle/dT$.
We also obtain the momentum integrated spectral function $A(\omega)$,
the analog of the density of states in the presence of interactions,
by analytic continuation of the Green function
via the maximum entropy method~\cite{gubernatis91,jarrell96}.

\begin{figure}[t]
\includegraphics[scale=0.35]{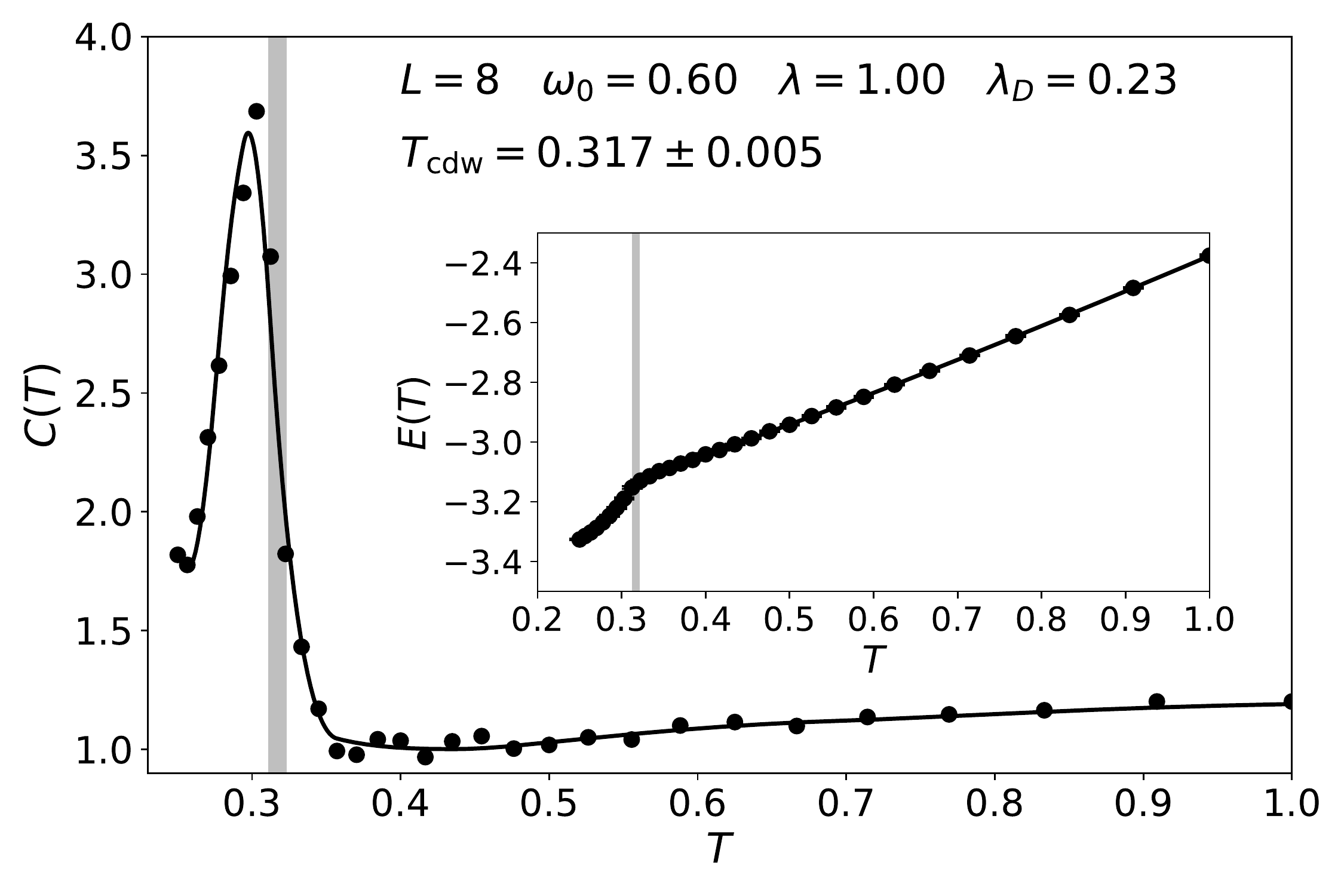}
\caption{
Specific heat $C(T)$ as a function of temperature $T$.
The low temperature peak corresponds to the onset of charge ordering.
Here $\lambda_D=0.23, \omega_0=0.60$ and the lattice size
is $N=8^3$.
}
\label{fig:CT}
\end{figure}

\noindent
\underbar{Correlation Length and Charge Structure Factor.}
At half-filling on a bipartite lattice the formation of a CDW phase is the fundamental ordering tendency of the Holstein model.  
At intermediate temperatures we observe the
formation of 
local pairs due to the effective 
on-site attraction $U_{\rm eff} = -\lambda^2/\omega_0^2$, 
between up and down electrons.
At lower $T$, the positions of the
pairs become correlated, since the lowering of energy
by virtual hopping is maximized by $-4t^2/U_{\rm eff}$ 
if each pair is surrounded by empty sites. A clear signature of this low temperature physics is seen in the heat capacity $C(T)$ as the temperature is lowered, which has a sharp peak at $T \sim 0.28$ corresponding to the CDW phase transition, as shown in Fig.~\ref{fig:CT}.


\begin{figure}[t]
\includegraphics[scale=0.35]{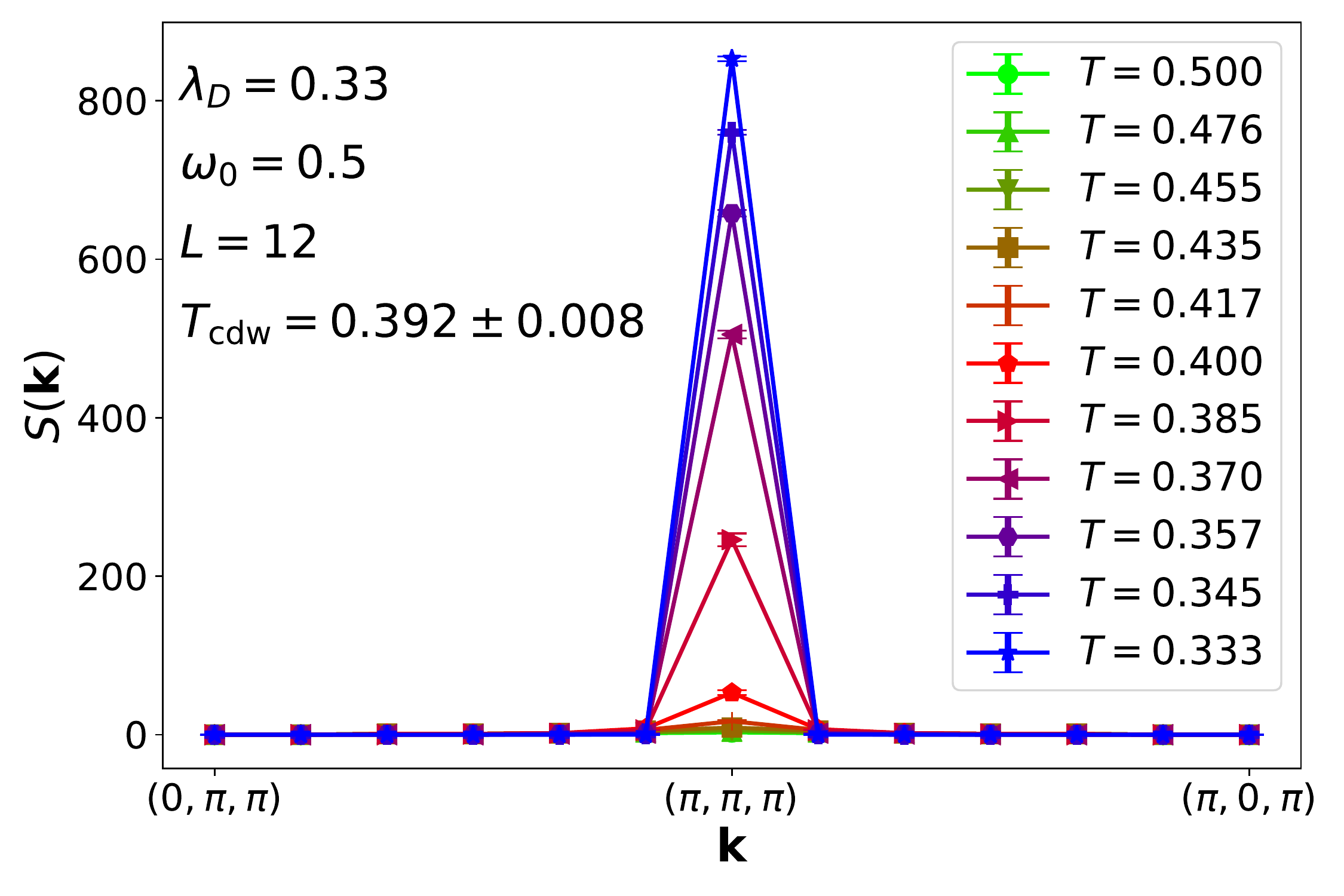}
\caption{Charge structure factor as a function of momentum for different 
inverse temperature $\beta$ at fixed
$\lambda_D=0.33$ and $\omega_0=0.5$.  As $T$ decreases, a peak develops
at ${\bf k}=(\pi,\pi,\pi)$.
The most rapid growth is for $T \sim 0.37$-$0.40$.
Finite size scaling analysis of the crossings of $S_{\rm cdw}$ in Fig.~\ref{fig:FSS}, precisely identifies
$T_c \sim 0.392 \pm 0.008$.}
\label{fig:Sk_vs_k}
\end{figure}

\begin{figure}[t]
\includegraphics[scale=0.35]{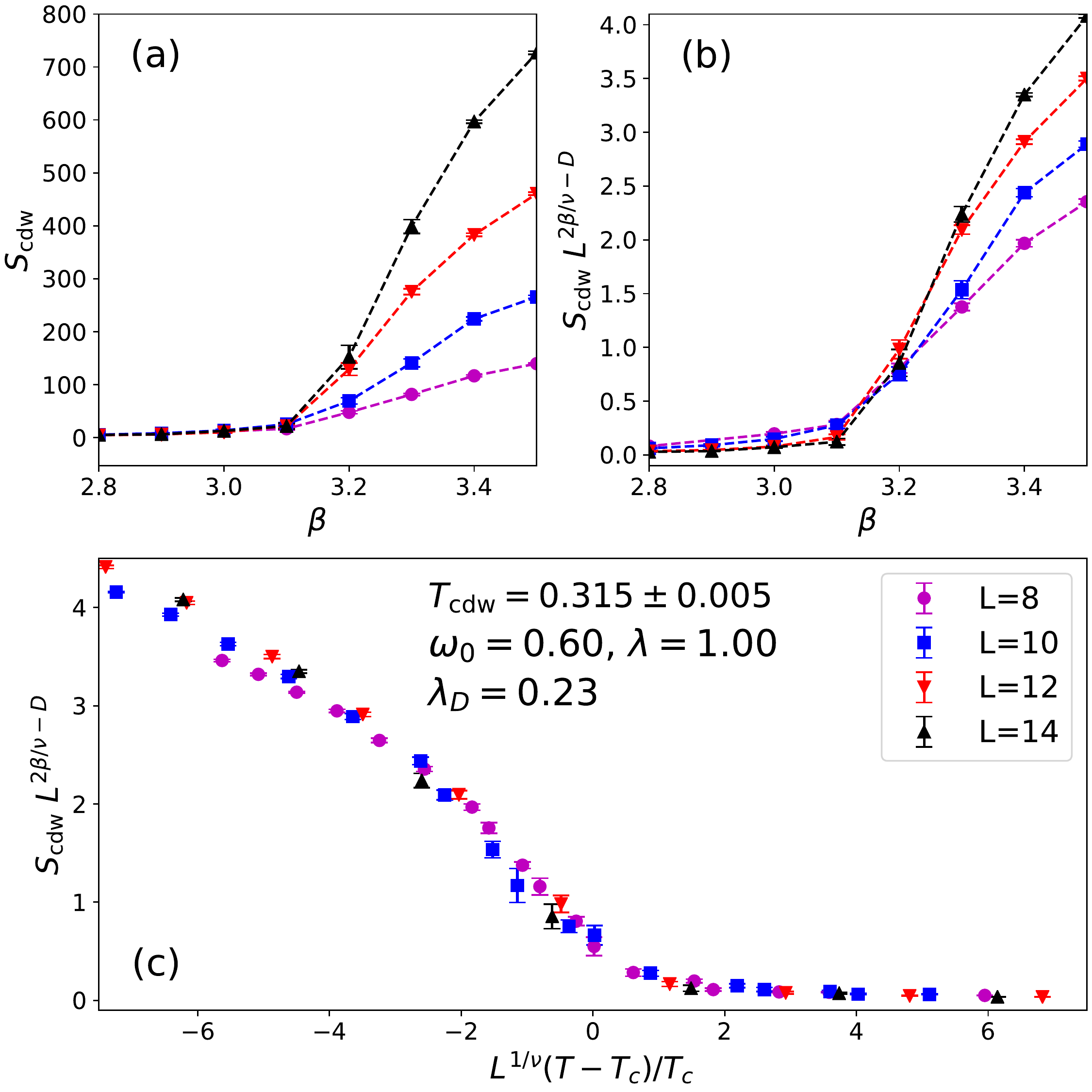}
\caption{
Finite size scaling analysis of the CDW structure factor.  
Panel (a) contains the raw (unscaled) data.  $S_{\rm
cdw}$ is independent of $L$ for small $\beta$ where the correlation
length is short.  At large $\beta$, $S_{\rm cdw}$ grows with $L$. 
Panel (b) scales $S_{\rm cdw}$ only.  The result
is a crossing plot which yields the critical inverse temperature
$\beta_c\, t = 3.15 \pm 0.05$.
The main panel (c) shows a full scaling plot where the data collapse
in a range of inverse temperatures near the critical point.
Holstein model parameters are
$\omega_0=0.60, \lambda=1.0$ so that $\lambda_D=0.23$.
}
\label{fig:FSS}
\end{figure}

It is also possible to detect the formation of this low temperature CDW phase by studying the density-density correlation function, and its Fourier transform, the charge structure factor, $S({\bf k})$. In Fig.~\ref{fig:Sk_vs_k} we show $S({\bf k})$, Eq.~\eqref{eq:Sandf}, versus ${\bf k}$
for different $T=\beta^{-1}$ and $\lambda_D=0.33 \ (\omega_0=0.5, \ \lambda=1.0)$.  
We see that, as $T$ is lowered, the peak height at ${\bf k}
= (\pi,\pi,\pi)$ increases by two orders of magnitude.
The value of $\beta$ for which the height increases most rapidly
provides a rough value for the transition temperature,
which can be more precisely determined via finite size scaling
(Fig.~\ref{fig:FSS}).

\color{black}
In real space, the density-density correlation function exhibits
a pattern which oscillates in sign on the two sublattices,
consistent with dominant ordering at ${\bf k}=(\pi,\pi,\pi)$
seen in Fig.~\ref{fig:Sk_vs_k}.  Above $T_c$, the
correlations die off exponentially, with a correlation
length $\xi$ which grows as $T \rightarrow T_c$.  
(See Supplemental Materials.)
In finite size simulations, $\xi$ will be bounded by the system size $L$, but one can nevertheless estimate it via~\cite{sandvik10},
\begin{align}
\xi = \frac{L}{2\pi} \sqrt{ 
\frac{S(q_1)/S(q_2)-1}
{4-S(q_1)/S(q_2)} } \,\,,
\label{eq:xi}
\end{align}
where $q_1=(\pi,\pi,\pi-\frac{2\pi}{L})$ and
$q_2=(\pi,\pi,\pi-\frac{4\pi}{L})$ are the two closest
wave vectors to the ordering vector
${\bf k}=(\pi,\pi,\pi)$.

\begin{figure}[t]
\includegraphics[height=6.0cm,width=8.0cm]{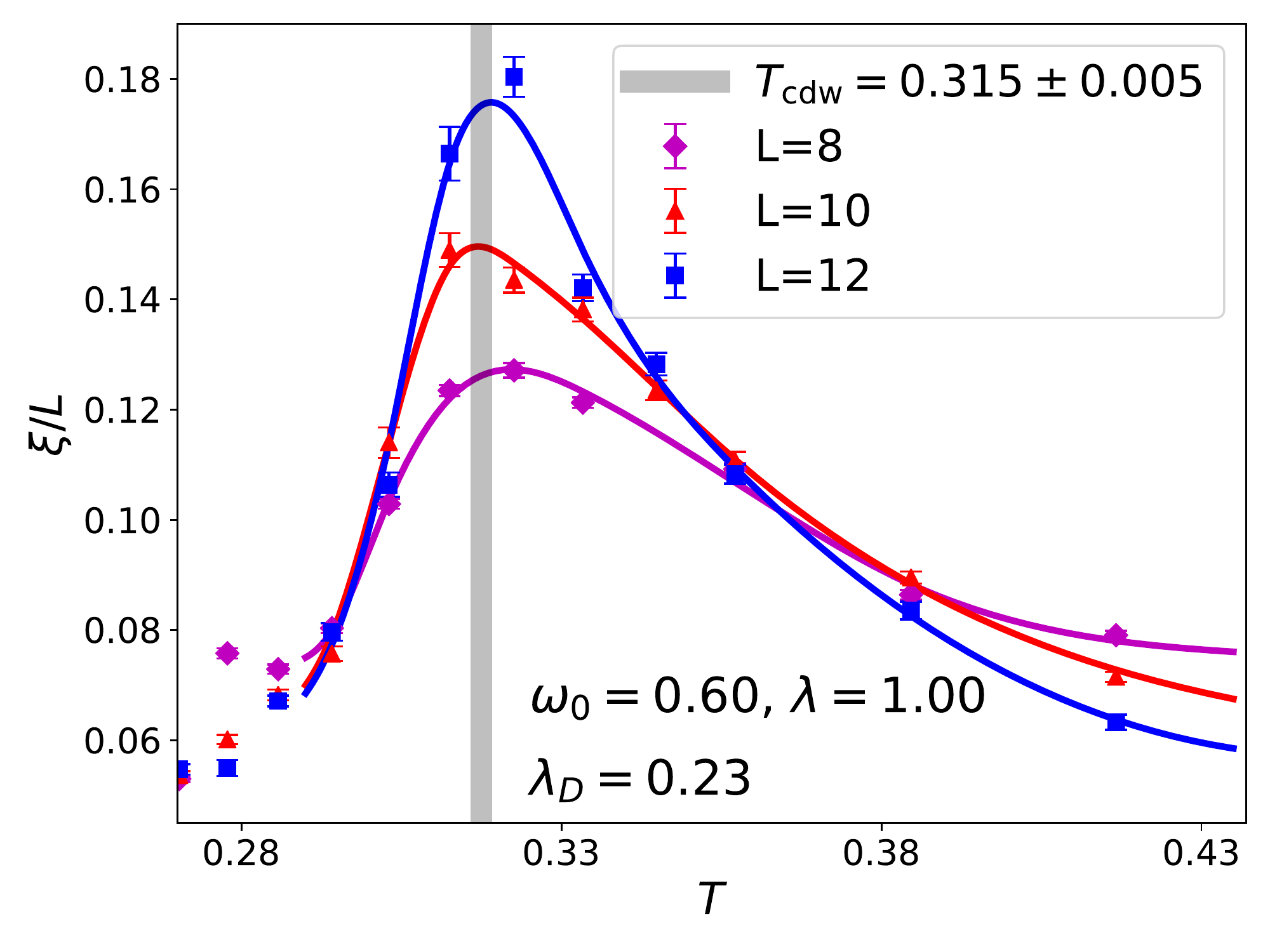}
\caption{
Correlation length obtained from Eq.~\eqref{eq:xi} with $\omega_0=0.6,
\lambda=1.0  \,\,(\lambda_D=0.23)$.
Shaded gray bar shows the value of $T_c$ obtained
from a finite-size scaling analysis of
the CDW structure factor (Fig.~\ref{fig:FSS}).
}
\label{fig:xi}
\end{figure}

Figure~\ref{fig:xi} shows the ratio $\xi/L$ as a function of temperature
for three lattice sizes $L=8,10,12$.
$\xi/L$ exhibits a characteristic peak, which sharpens
with increasing lattice size.  
In the following section, we will present data indicating $T_{\rm
cdw}=0.31$ which is consistent with the peak in finite lattice
sizes approaching $T_c$ from above in our data as well.

\noindent
\underbar{CDW Transition.}
Having seen the essential qualitative effects of the
electron-phonon coupling, we now perform finite size scaling to
locate the transition precisely.
The three panels of Fig.~\ref{fig:FSS} exhibit the steps in this
process.  The upper left panel (a) exhibits raw data for $S_{\rm cdw}$
versus inverse temperature $\beta$.  At high $T$ (small $\beta$)
the values of $S_{\rm cdw}$ for different system sizes coincide with
each other, because the charge correlations are short ranged and the
additional large distance values in the 
sum over ${\bf r}$ in Eq.~\eqref{eq:Sandf},
present as $L$ increases, make no contribution.
However, as $T$ decreases ($\beta$ increases) the correlation 
length reaches the lattice size, and values of $S_{\rm cdw}$ 
now become sensitive to the cut-off $L$.
As a consequence, a crude estimate of $T_{\rm cdw}$ can already be made
as the temperature at which the curves begin to separate,
i.e.~$T_{\rm cdw} \sim 0.31 \; (\beta_{c} \sim 3.2)$.

A much more accurate determination of $T_{\rm cdw}$ is provided
by making a crossing plot 
(Fig.~\ref{fig:FSS}c) 
of $S_{\rm cdw} L^{2\beta/\nu-D}$
versus $\beta$.  Curves for different lattice sizes $L$ should cross at
$\beta_c=1/T_{\rm cdw}$.  In this analysis we make use of the expected
universality class of the transition, the 3D Ising model, to 
provide values for the exponents 
$\beta=0.326$ and
$\nu=0.63$.
We conclude $T_{\rm cdw}= 0.315 \pm 0.005$.
Finally, Fig.~\ref{fig:FSS}(c)
gives the full scaling collapse,
using $T_{\rm cdw}$ from panel (b) and again
employing 3D Ising exponents.


\begin{figure}[t]
\includegraphics[height=6.0cm,width=8.0cm]{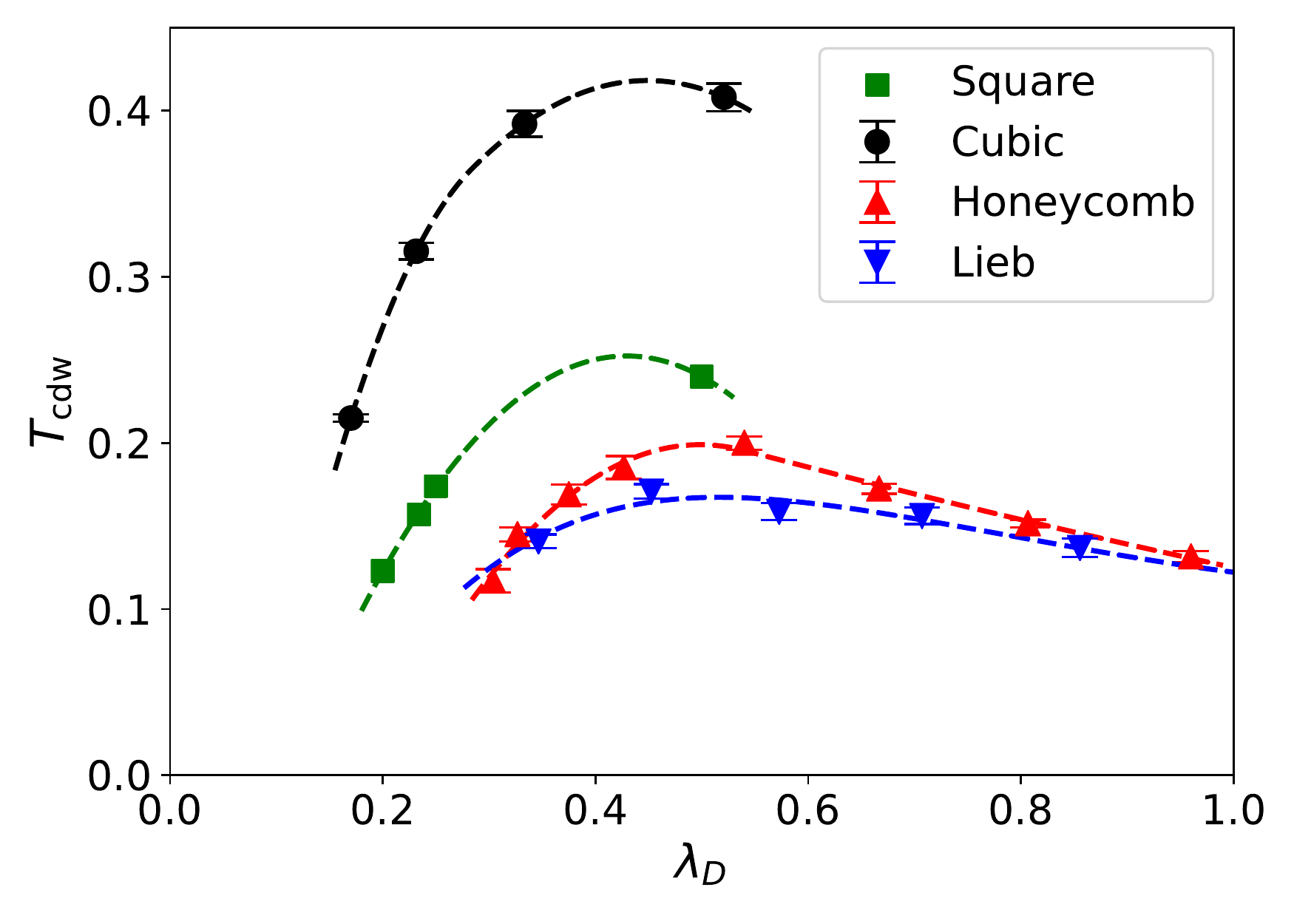}
\caption{Phase diagram of the 3D Holstein model on a cubic lattice as a function of $\lambda_D$, with $\lambda=1$ held fixed.
For comparison, critical temperatures on three 2D lattice
geometries, square, honeycomb, and Lieb are also given.
}
\label{fig:phase_diagram}
\end{figure}

Combining plots like those of Fig.~\ref{fig:FSS} for different
values of $\lambda$ and $\omega_0$ allows us to obtain the
finite temperature phase diagram of the 3D Holstein model, 
Fig.~\ref{fig:phase_diagram}, which is the central result of this paper.
We see that $T_c$ is increased by roughly a factor of two in
going from various 2D geometries (square~\cite{weber18}, 
Lieb~\cite{feng20}, and honeycomb~\cite{chen19,zhang19})
to 3D.  This increase is quite similar to that of going from 
2D square ($T_c \sim 2.27$) to 3D cubic ($T_c \sim 4.51$) 
for the CDW transition of {\it classical} lattice gas (Ising) model.

\begin{figure}[t]
\includegraphics[scale=0.35]{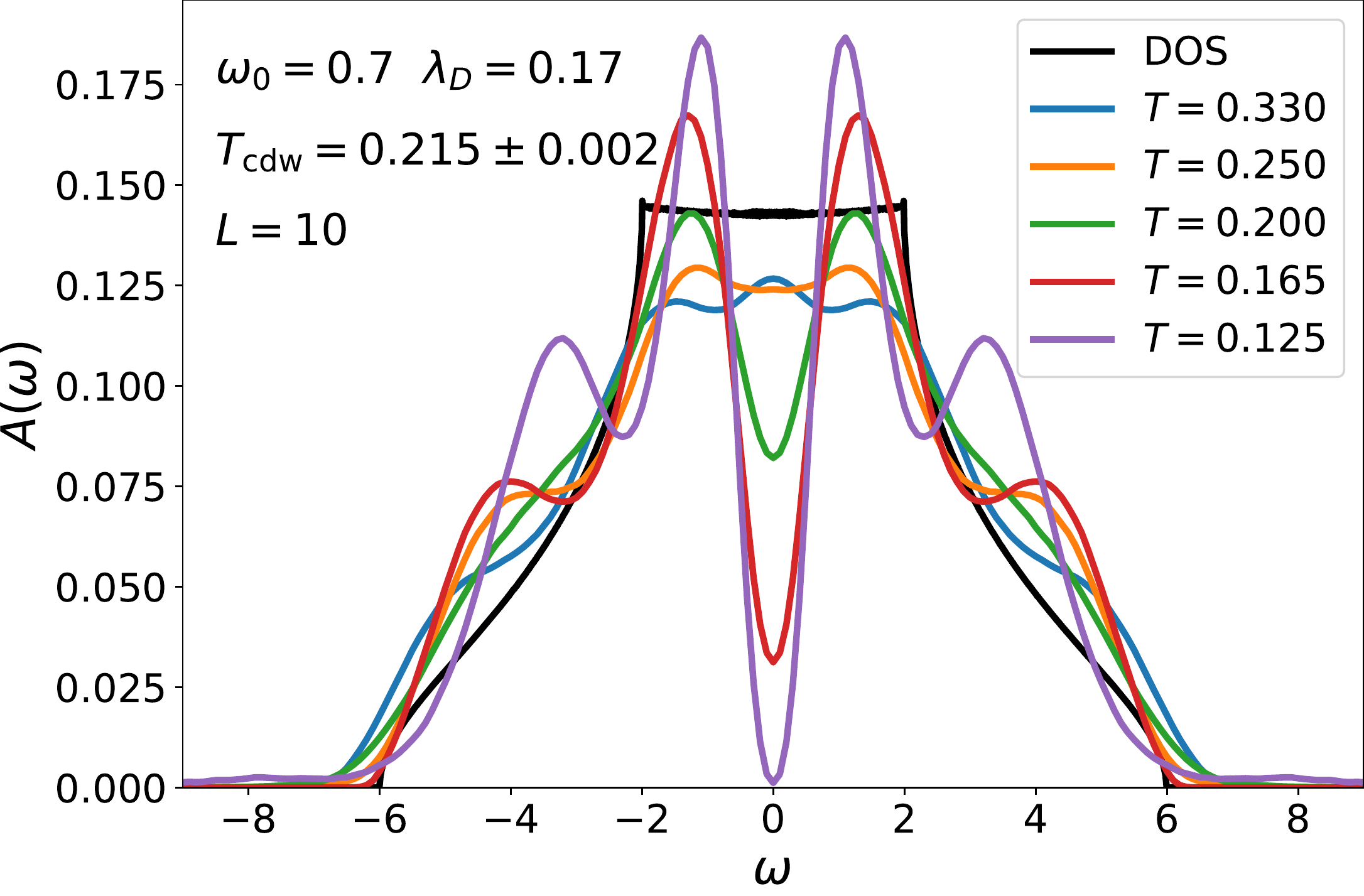}
\caption{
Momentum integrated spectral function $A(\omega)$.
Here $\omega_0=0.7, \lambda_D=0.17$,
and the lattice size $N=10^3$.
A suppression of $A(\omega=0)$ coincides with 
reaching $\beta_c \sim 5$.  
(See Fig.~\ref{fig:phase_diagram}.)
A full gap develops at a somewhat lower temperature.
Also shown, for comparison, is the density of states
of non-interacting electrons ($\lambda_D=0$) hopping on a cubic lattice.
}
\label{fig:Aw}
\end{figure}

\noindent
\underbar{Spectral Function.}
The preceding results are all obtained with imaginary 
time-{\it independent} Green functions.  More generally, one can
consider,
\begin{align}
G({\bf k},\tau) \equiv
\langle  c({\bf k},\tau)
 c^{\dagger}({\bf k},0)  \rangle
= \int d\omega 
A({\bf k},\omega)  \,
\frac{e^{-\omega \tau}}{e^{\beta \omega}+1}
\label{eq:Aw}
\end{align}
to determine the spectral function $A({\bf k},\omega)$.
This involves inverting the integral relation in
Eq.~\eqref{eq:Aw} using analytic continuation~\cite{gubernatis91,jarrell96}.
This is the first use of our Langevin approach for dynamical
behavior.  
Figure~\ref{fig:Aw}
shows $A(\omega)$ for several different temperatures at
fixed $\omega_0=0.7, \lambda_D=0.17$.
At high temperatures ($\beta=3$ and $4$) 
the main effect of the electron-phonon 
interaction is to increase the spectral function somewhat in the region
close to the band edges $\omega = \pm 6t$.
The renormalized bandwidth is remarkably unchanged
from that of free electrons on a cubic lattice, $W=12t$.
When $T$ reaches the CDW ordering temperature, $\beta \sim 5$
(see Fig.~\ref{fig:phase_diagram}) $A(\omega=0)$ develops
a pronounced dip.  This suppression continues to increase
until, at $\beta=8$, $A(\omega=0)$ vanishes.
This sequence, in which a dip first signals entry into the
CDW phase, is consistent with the trends reported in~\cite{cohenstead19}.

\noindent
\underbar{Conclusions.}
We have used a new Langevin QMC method to study the Holstein
Hamiltonian on a three-dimensional cubic lattice.  This new approach allows us to access much larger lattice sizes, enabling us to perform a
reliable finite size scaling analysis to determine the CDW transition temperature. Using this method, we obtained results that, in momentum space, were sufficient to resolve the width
of the charge structure factor peak and the smearing of the Fermi
surface by electron-phonon interactions.  The specific heat and
spectral function provide useful alternate means to examine
the low temperature properties.  Their behavior is consistent with
that seen by direct observation of charge correlations.

While a single band model of interacting electrons
does seem to provide 
a reasonably accurate representation of cuprate physics~\cite{scalapino94} 
(although not that of the iron-pnictides), realistic
CDW materials generally have much richer band structures.
Since, at a formal level, additional sites and additional
orbitals are equivalent in real-space QMC simulations,
an ability to simulate larger spatial lattices also opens
the door to the study of more complex CDW systems.
Of course, the accurate description of these materials
requires not only several electronic bands, but also a refinement
of the description of the phonons and electron-phonon coupling,
which are also treated at a very simple level in the Holstein
Hamiltonian.  Initial steps to include phonon dispersion
have recently been made~\cite{costa18}.  However,
refinements to the electron-phonon coupling such
as a momentum dependent $\lambda({\bf q})$ remain
a challenge to simulations because of the phase separation
that results in the absence of electron-electron repulsion~\cite{xiao19}.


\vskip0.03in \noindent \underbar{Acknowledgements:}  
The work of B.C-S.~and R.T.S~was supported by the grant DE‐SC0014671 funded by
the U.S. Department of Energy, Office of Science.
K.B. acknowledges support from the center of Materials Theory as a part of the Computational Materials Science (CMS) program, funded by the U.S. Department of Energy, Office of Science.
G.G.B.~is partially supported by the French government, through the
UCAJEDI Investments in the Future project managed by the National
Research Agency (ANR) with the reference number ANR-15-IDEX-01.
C.C.~and Z.Y.M.~acknowledge the supports from MOST China through the National Key Research and Development
Program (Grant No.~2016YFA0300502) and Research Grants Council of Hong Kong SAR China through 17303019 and thank the Center for Quantum Simulation Sciences in the Institute of Physics, Chinese Academy of Sciences, the Computational Initiative at the Faculty of Science at the University of Hong Kong, the Platform for
Data-Driven Computational Materials Discovery at the Songshan Lake Materials Laboratory, Guangdong, China and the National Supercomputer Centers in Tianjin and Guangzhou for their
technical support and generous allocation of CPU time.


\bibliography{cubicholstein}

\end{document}


\title{Supplemental Materials:
Langevin Simulations of the Half-Filled Cubic Holstein Model}
\author{Ben Cohen-Stead} 
\affiliation{Department of Physics, University of California, 
Davis, CA 95616, USA}
\author{Kipton Barros}
\affiliation{Los Alamos National Laboratory, Physics and Chemistry of
Materials, Theoretical Division, Los Alamos, 87545}
\author{Z.Y. Meng}
\affiliation{Beijing National Laboratory for Condensed Matter Physics
and Institute of Physics, Chinese Academy of Sciences, Beijing 100190,
China}
\affiliation{Department of Physics and HKU-UCAS Joint Institute of Theoretical and Computational Physics, The University of Hong Kong, Pokfulam Road, Hong Kong, China}
\affiliation{Songshan Lake Materials Laboratory, Dongguan, Guangdong 523808, China}
\author{Chuang Chen}
\affiliation{Beijing National Laboratory for Condensed Matter Physics
and Institute of Physics, Chinese Academy of Sciences, Beijing 100190,
China}
\author{R.T. Scalettar}
\affiliation{Department of Physics, University of California, 
Davis, CA 95616, USA}
\author{George Batrouni}
\affiliation{Universit\'e C\^ote d'Azur, INPHYNI, CNRS, 0600 Nice,
France}
\affiliation{MajuLab, CNRS-UCA-SU-NUS-NTU International Joint Research
  Unit, 117542 Singapore}
\affiliation{Centre for Quantum Technologies, National University of
  Singapore, 2 Science Drive 3, 117542 Singapore}
\affiliation{Department of Physics, National University of Singapore, 2
  Science Drive 3, 117542 Singapore}
\affiliation{Beijing Computational Science Research Center, Beijing,
100193, China}

\maketitle

Here we provide additional details concerning the
charge order in the half-filled cubic Holstein model, to wit
the individual components of the energy,
real space density-density
correlations, and mean field phase diagram.

\section{1.  Energy Components}

\begin{figure}[h!]
\includegraphics[scale=0.37]{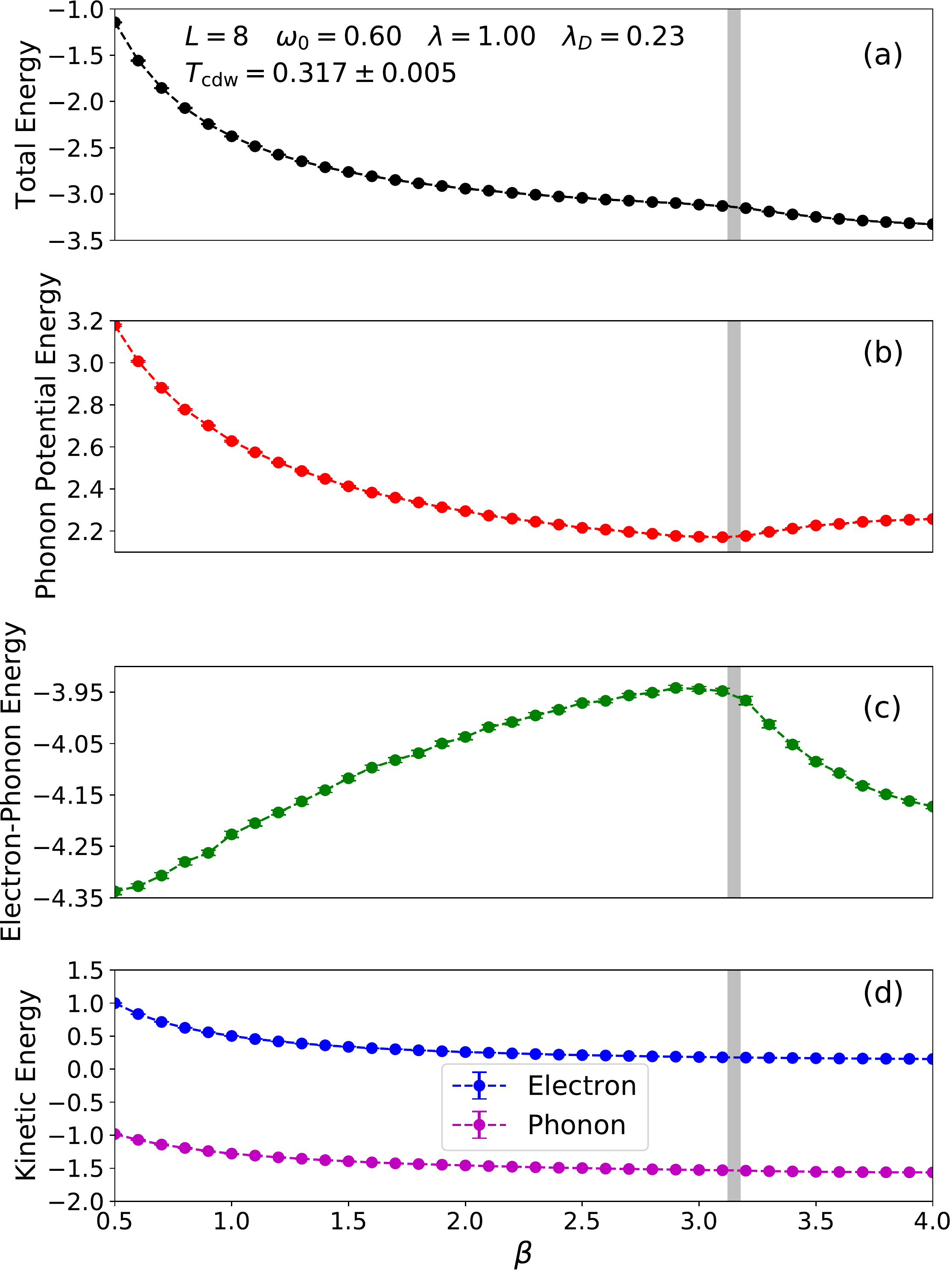}
\caption{
Total energy (panel a) and 
individual components of the energy (panels b,c,d) of
the half-filled cubic Holstein model for $\omega_0/t=0.6$
and $\lambda_D=0.23$.
The phonon potential energy (panel b) is non-monotonic,
with a weak minimum at $T_c$.
The electron-phonon energy (panel c)
shows the sharpest signature of the CDW transition.
The electron and phonon kinetic energies (panel d) show
little evidence of the transition.
}
\label{fig:SMCT}
\end{figure}

The inset to Fig.~1 of the text gave the temperature evolution
of the {\it total energy}, with the main panel showing its derivative,
the specific heat $C(T)$.  A sharp peak at low temperature, $T/t \sim 0.30$,
signals the CDW transition.  Figure \ref{fig:SMCT}
gives the {\it individual} components of the energy.
The phonon potential energy and electron-phonon energies exhibit
clear signatures through $T_{\rm cdw}$, with the former having a minimum,
and the latter a maximum there.
The electron-phonon energy initially rises as $\beta$ increases, 
i.e.~contributing negatively to the specific heat, $dE_{\rm elph}/dT < 0$, 
and then abruptly drops beyond $\beta_{\rm cdw}$.

\section{2.  Real Space Density Correlations}

\begin{figure}[h!]
\includegraphics[height=6.0cm,width=8.0cm]{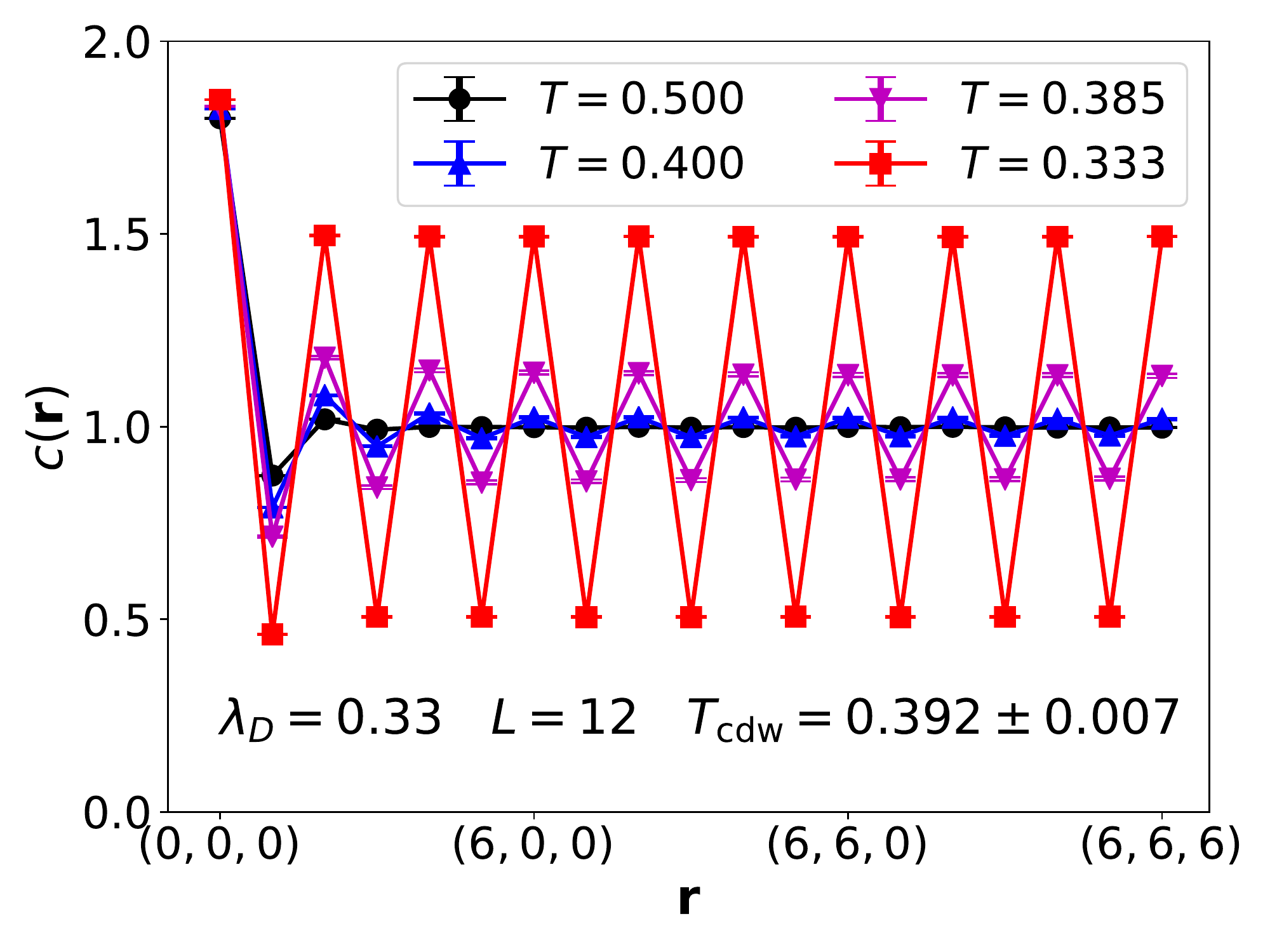}
\caption{
Density-density correlations versus separation for the half-filled
cubic Holstein model with frequency $\omega_0=0.5$,
electron-phonon coupling $\lambda_D = 0.33$ on a $12^3$ lattice.
Oscillations in $c({\bf r})$ are barely observable
at $T=0.400$, but clearly present at $T=0.385$.
}
\label{fig:SMdenden}
\end{figure}

The analysis in the main text focused on the (momentum space) 
charge structure factor $S({\bf k})$
since it most clearly captures the CDW phase transition temperature.
However, it is also useful to observe the density
correlations, since they exhibit the long range 
spatial order.  Figure \ref{fig:SMdenden}
shows $c({\bf r})= \langle \, n_{\bf j+r} \, n_{\bf j} \, \rangle$
along a trajectory of steadily increasing separation
${\bf r} = (0,0,0)$ to $(6,0,0)$ to $(6,6,0)$, and
finally, to $(6,6,6)$, the last being
the maximal attainable separation on a $12^3$ lattice with periodic
boundary conditions.
For high temperatures
$c({\bf r})= \langle \, n_{\bf j+r} \,\rangle \, 
\langle \, n_{\bf j} \, \rangle = 1$, its uncorrelated value, once ${\bf r}$ is beyond a few lattice spacings.
In contrast,
a clear pattern of distinct $c({\bf r})$ on
the two sublattices is present at low temperatures.
Very little spatial decay is seen in $c({\bf r})$ at low $T$.
The temperature at which this pattern
emerges is consistent with $T_c$ from Fig.~5 of the
main text, obtained by finite size scaling of $S({\bf k})$.

\section{3.  Mean Field Theory}

The failures of Mean Field Theory (MFT) for classical phase transitions
are well known-  an overestimation of the tendency to order, e.g.~critical
temperatures which are substantially
greater than the exact values, and, of course,
incorrect scaling exponents when the dimension is less than the 
upper critical dimension.  For itinerant Fermi systems, MFT
has an additional weakness:  it is unable to distinguish the
formation temperatures $T_*$ of local moments (in the case of magnetism)
or doubly occupied sites (in the case of charge density wave order),
from the temperatures at which these local objects achieve
regular long range patterns, i.e.~the transition temperature $T_c$.
Thus, for antiferromagnetism in the half-filled Hubbard model,
$T_{\rm af}^{\rm mft} \sim U$
whereas $T_{\rm af}^{\rm exact} \sim J = 4t^2/U$.  
As a consequence, at large coupling $U$, $T_{\rm af}$ is especially 
poorly captured and indeed has a fundamentally erroneous dependence
on $U$.

\begin{figure}[h!]
\includegraphics[scale=0.35]{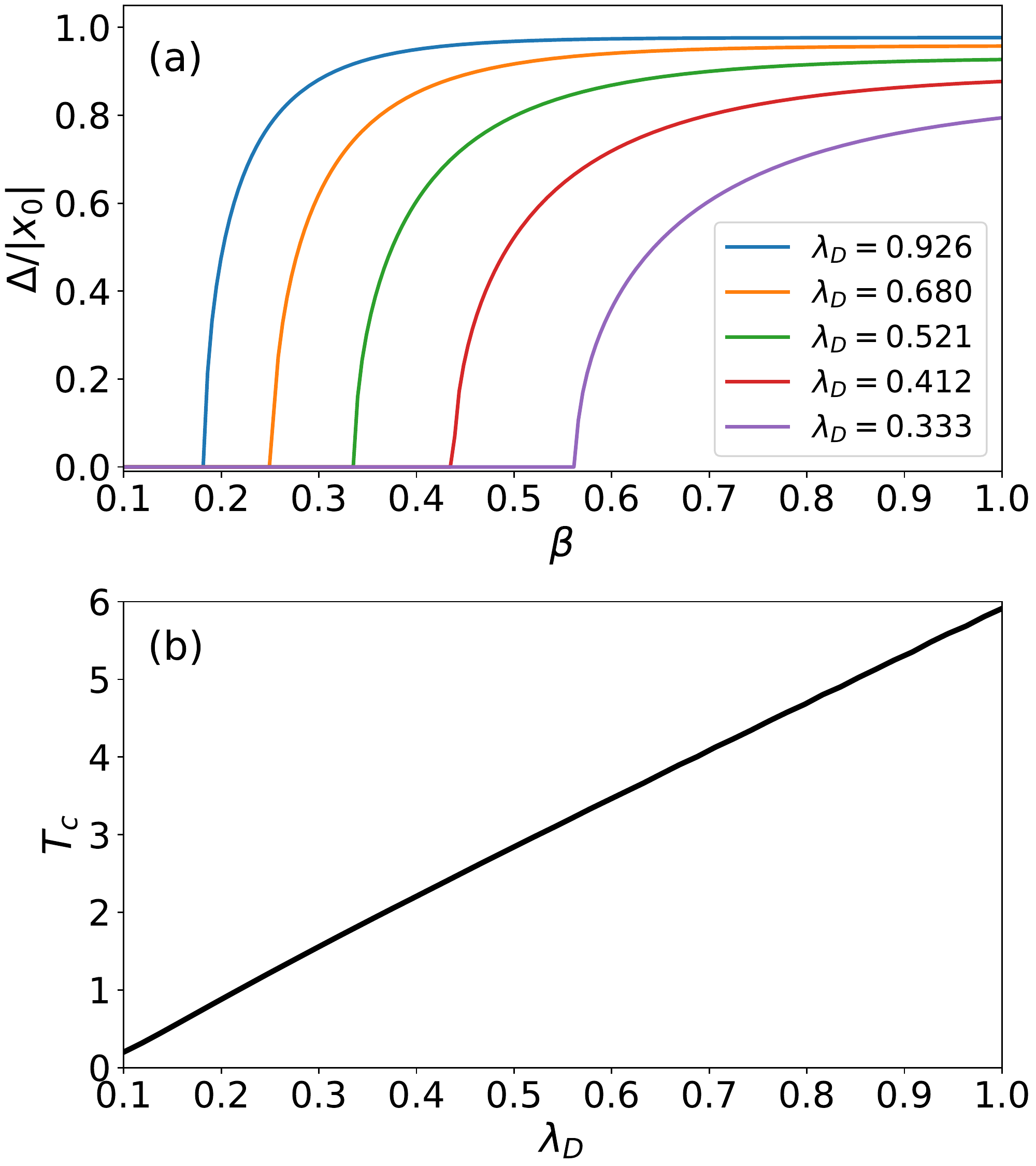}
\caption{
Panel (a) shows the order parameter as a function of inverse temperature $\beta$ for different $\lambda_D$.
Panel (b) shows the CDW transition temperature given by mean-field theory for the cubic Holstein model at
half-filling as a function of dimensionless coupling $\lambda_D$
at (with $\lambda=1$ held fixed).
$T_c$ increases linearly with $\lambda_D$ at strong coupling, in
contrast to the non-monotonic behavior of the QMC results of
Fig.~5 of the main text.
At small coupling the MFT becomes difficut to converge, hence $T_c$ is
not shown for $\lambda_D \lesssim 0.1$.
}
\label{fig:SMmft}
\end{figure}

The same is true for the CDW transition in the half-filled 3D Holstein model.
Figure \ref{fig:SMmft} shows $T_c$ within MFT.
The exact $T_c$ (Fig.~5 of main text) reaches a maximal value of
$T_c/t \sim 0.4$ for $\lambda_D \sim 0.42$.  
The MFT transition temperature at this $\lambda_D$ is overestimated
by a factor of about five.  This is a much greater difference than
between the transition temperature of charge ordering in a
classical lattice gas, where the MFT $T_c$ is only a factor of about 4/3
larger than the exact (monte carlo) result.

As for the Hubbard model, MFT will not only overestimate the transition
temperature for ordering at half-filling, but will also greatly
enlarge
the region of densities about half-filling at which 
charge order occurs\cite{hirsch85}.

\bibliography{supmat.bib}